\documentclass[prl, showpacs, preprintnumbers, amsmath, amssymb, superscriptaddress, twocolumn]{revtex4}
\usepackage{graphicx, color, dcolumn, bm, natbib, amssymb, amsmath}
\usepackage[normalem]{ulem}
\begin{document}

\title{The Dzyaloshinskii-Moriya interaction is under control: an orchestrated flip of the chiral link between structure and magnetism for Fe$_{1-x}$Co$_x$Si}

\author{S.-A. Siegfried}
\affiliation{Helmholtz Zentrum Geesthacht, Geesthacht, 21502, Germany}
\author{E. V. Altynbaev}
\affiliation{Petersburg Nuclear Physics Institute, Gatchina, St Petersburg, 188300, Russia}
\affiliation{Saint-Petersburg State University, Ulyanovskaya 1, Saint-Petersburg, 198504, Russia}
\author{N. M. Chubova}
\affiliation{Petersburg Nuclear Physics Institute, Gatchina, St Petersburg, 188300, Russia}
\author{V. Dyadkin}
\affiliation{Swiss-Norwegian Beamlines at the European Synchrotron Radiation Facility, Grenoble, 38000 France}
\affiliation{Petersburg Nuclear Physics Institute, Gatchina, St Petersburg, 188300, Russia}
\author{D. Chernyshov}
\affiliation{Swiss-Norwegian Beamlines at the European Synchrotron Radiation Facility, Grenoble, 38000 France}
\author{E. V. Moskvin}
\affiliation{Petersburg Nuclear Physics Institute, Gatchina, St Petersburg, 188300, Russia}
\affiliation{Saint-Petersburg State University, Ulyanovskaya 1, Saint-Petersburg, 198504, Russia}
\author{D. Menzel}
\affiliation{Institut f\"ur Physik der Kondensierten Materie, TU Braunschweig, 38106 Braunschweig, Germany}
\author{A. Heinemann}
\affiliation{Helmholtz Zentrum Geesthacht, Geesthacht, 21502, Germany}
\author{A. Schreyer}
\affiliation{Helmholtz Zentrum Geesthacht, Geesthacht, 21502, Germany}
\author{S. V. Grigoriev}
\affiliation{Petersburg Nuclear Physics Institute, Gatchina, St Petersburg, 188300, Russia}
\affiliation{Saint-Petersburg State University, Ulyanovskaya 1, Saint-Petersburg, 198504, Russia}

\date{\today}

\begin{abstract}

Monosilicides of 3d-metals frequently show a chiral magnetic ordering with the absolute configuration defined by the  chirality of the crystal structure and the sign of the Dzyaloshinskii-Moriya interaction (DMI). Structural and magnetic chiralities are probed here for Fe$_{1-x}$Co$_x$Si series and their mutual relationship is found to be dependent on the chemical composition. The chirality of crystal structure was previously shown to be governed by crystal growth, and the value of the DMI is nearly the same for all monosilicides of Fe, Co and Mn. Our findings indicate that the sign of the DMI in Fe$_{1-x}$Co$_x$Si is controlled by the Co composition $x$. We have been able to directly measure the change of the link between structure and magnetism in this helimagnetic B20 alloy.

\end{abstract}

\maketitle

Scattering of polarized neutrons on chiral magnetic structures allows one to determine the absolute magnetic configuration, thus left- and right-handed helices can be easily distinguished \cite{Maleyev1995PRL}. On the other hand, knowing the magnetic configuration, one can analyse the polarization of a scattering beam \cite{Chubova_Surface_Poverhnost2014}. Similar effects could also help to manipulate spin polarization of an electron current providing that the electrons interact with the known chiral magnetic structure.

The ability to manipulate the electron spin is a necessary component for the spintronics \cite{SinovaZutic_NatMat_2012}, thus magnetic chiral organic molecules \cite{RNaaman_JPhysChemLett_v3_p2178_2012} or large scale magnetic structures have been proposed as such tools \cite{Mochizuki_NatMatt_v13_p241_2014}.
However, the question how to get the magnetic structure of a necessary chirality for spintronics applications is still open. Here we address the question for the case of Fe$_{1-x}$Co$_x$Si solid solutions which, for certain compositions, show chiral (spiral) magnetic ordering \cite{Dzyaloshinskii64ZETF,BakJPCM80_v13,Kataoka_SSC_1980}.

The structural chirality in monosilicides of 3d-metals is solely controlled by crystal growth \cite{Dyadkin_PRB_2011}.
A link between the structural and magnetic chiralities is provided by the Dzyaloshinskii-Moriya interaction (DMI) and has been experimentally proved for many monosilicides of 3d-metals \cite{Grigoriev_PRB_2010,Dyadkin_PRB_2011,Grigoriev_PRL_2013, Morikawa_PRB_2013}. 
The strength of the DMI defines the pitch of the magnetic spiral while the sign of the DMI sets a relationship between structural and magnetic chiralities to be the same or opposite \cite{BakJPCM80_v13,Dmitriev_JPhysCondMatt_2012}.
%However, the microscopic origin of the DMI is not yet clear, and both strength and sign of the DMI can hardly be predicted from a theory. Here we show a phenomenological route towards a control of the chiral magnetism.

For powder samples of Mn$_{1-x}$Fe$_x$Ge \cite{Grigoriev_PRL_2013,Tokura_2} and Fe$_{1-x}$Co$_x$Ge \cite{Grigoriev_PRB_2014} it was shown that the spiral wave vector $k = 2\pi/d$, where $d$ is the spiral period, goes to zero value at a certain composition. 
The monotonic behavior of the wave vector indicates that the DMI goes to zero at the very same composition and, therefore, should change its sign as a function of $x$ \cite{Grigoriev_PRL_2013,Grigoriev_PRB_2014}.
Recently, it was possible to reproduce the observed change of {\it D} in Mn$_{1-x}$Fe$_x$Ge using density-functional theory calculations \cite{Koretsune_arxiv_2015,Gayles_arxiv_2015}. The dynamic of the $d_{x^2-y^2}$-like states has been revealed as the main mechanism behind the change of the sign of {\it D}. With increasing $x$ in Mn$_{1-x}$Fe$_x$Ge the $d_{x^2-y^2}$-like states move from above to below the Fermi energy, become occupied and enter the region of $d_{xy}$-states with opposite spin, leading to the change of the sign of the DMI \cite{Gayles_arxiv_2015}.

Here we further exploit the idea to control the DMI sign for the monosilicide series Fe$_{1-x}$Co$_x$Si. 
At variance with the germanides, the silicides can be grown as single crystals with controlled structural chirality \cite{Dyadkin_PRB_2011}. A large size of crystals also makes possible a combined determination of the structural $\Gamma_c$ and magnetic $\gamma_m$ chiralities by resonant x-ray diffraction and polarized neutron scattering, correspondingly. Here and below chiralities are defined according to Ref. \cite{Dmitriev_JPhysCondMatt_2012}.
Taken together the two experimental probes allow us to follow $\Gamma_c\times\gamma_m$ as a function of composition $x$. Thus, the sign of the DMI term in the Hamiltonian of Ref. \cite{BakJPCM80_v13,Kataoka_SSC_1980} can be experimentally probed via the product $\Gamma_c\times\gamma_m$, allowing us to directly observe the flip of the link of the structural and magnetic chirality with the concentration $x$.

Single crystals of Fe$_{1-x}$Co$_x$Si were grown using the Czochralski technique for the following concentrations $x=0.5$, 0.6, 0.65, 0.7, 0.8. 
The same structural chirality of all grown crystals was provided by a consequent use of every grown crystal as the seed for the next one. As it was shown before this technique gives almost 100\% control of the structural chirality \cite{Dyadkin_PRB_2011}.
The absolute crystal structure can be established by the X-ray single crystal diffraction data providing that resonant contribution enables to observe violation of the Friedel law, more details can be found in \cite{Flack_1,Flack_2,Grigoriev_PRL_2009}.

Single crystal Bragg diffraction data were collected at the room temperature 
using the PILATUS@SNBL diffractometer at the BM01A end station of the 
Swiss-Norwegian Beamlines at the ESRF (Grenoble, France); the wavelength of the 
synchrotron radiation was set to 0.70135 \AA. The data were collected with a 
single $\phi$-scan with angular step of $0.1^\circ$ in a shutter-free mode with 
the Pilatus2M detector.
The raw data were preprocessed with SNBL Toolbox, the integral intensities 
were extracted from the frames with the CrysAlisPro software \cite{crys}, the 
crystal structure was solved with SHELXS and refined with SHELXL 
\cite{Seldrick_ActaCryst_1990}. 
Crystals with an average size of about 100 microns were cut from large single crystals.
The diffraction data summarized in Table \ref{tab:table1}. 

\begin{table}[h]
\caption{Diffraction data for Fe$_{1-x}$Co$_{x}$Si with $x = 0.6$, 0.65, 0.7, 0.8.}{\label{tab:table1}}
\begin{ruledtabular}
\begin{tabular}{l|ccc|cc|c}
$x$ & $R_i$ & $R_1$ & $R_w$ & $x_\mathrm{Me}$ & $x_\mathrm{Si}$  & Flack \\
\colrule 
0.6 & 0.025 &0.0295&0.0811 & 0.86000(2)&0.1579(5) & 0.02(5)\\
0.65 & 0.017&0.0110&0.0290 & 0.85989(9)&0.1572(2) & 0.01(4) \\
0.7 & 0.016&0.0299&0.0665 & 0.85968(19)&0.1575(4) & 0.10(7) \\
0.8 & 0.049&0.0193&0.0420 & 0.85882(13)&0.1571(3) & -0.01(7) \\
%0.9 & 0.024&0.033&0.081 & 0.8595(2)&0.1574(5) & 0.10(6)\\
\end{tabular}
\end{ruledtabular}
\end{table}

The data are of good quality and agreement with the structural  P2$_1$3 model is high as can be seen from R-factors. The unit cell dimensions follow the Veagard law but the atomic positions stay nearly the same as a function of composition; the absolute structure is defined according to their values. Thus, in agreement with definitions given in \cite{tanaka_JPSJ_1985, Grigoriev_PRL_2009}, the chirality $\Gamma_c$ of structure with $x_\mathrm{Me} \approx 0.86$ is set to $+1$.
The Flack parameter, which is a measure of presence of domains with the opposite chirality, is zero within $1 \div 2$ standard deviations; the results confirm the same absolute structure (i.e. the same structural chirality) for all the tested crystals, as expected from the crystal growth procedure.

\begin{figure}[ht]
        \begin{minipage}{0.99\linewidth}
        \center{\includegraphics[width=1\linewidth]{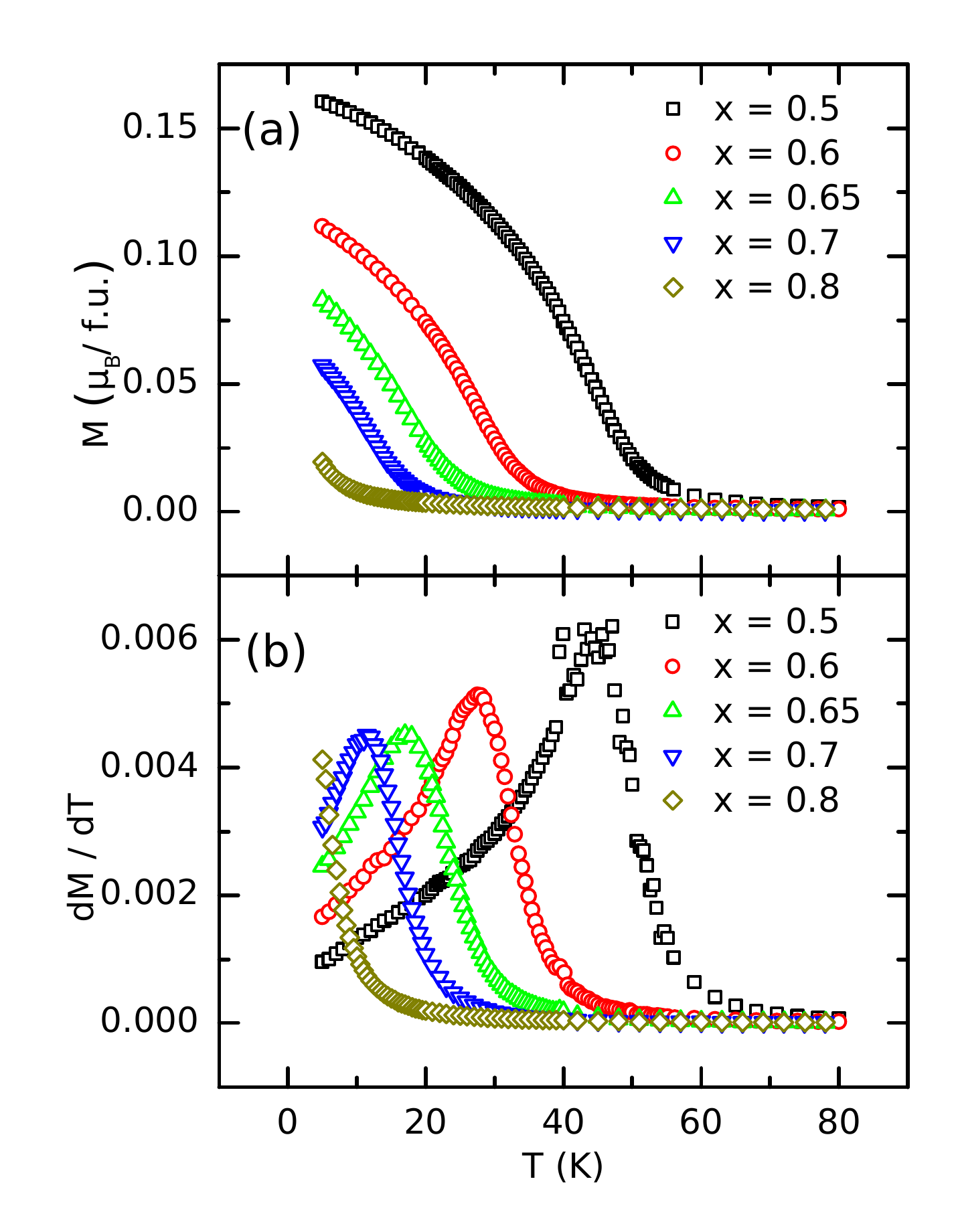}}
        \end{minipage}
        \caption{(color online). (a) The temperature dependence of the magnetization $M$ for Fe$_{1-x}$Co$_x$Si compounds with $x = 0.5 \div 0.8$ at $H = 100$ mT. (b) The first derivative of the magnetization on the temperature $dM/dT$.}
        \label{ris:M_vs_T_FeCoSi}
\end{figure}

Fe$_{1-x}$Co$_x$Si compounds are magnetically ordered in the concentration range $ 0.05 \leq x \leq 0.8$ \cite{Beille1,Beille2}. Magnetic measurements of newly synthesized samples were carried out with the SQUID-magnetometer Quantum Design MPMS\nobreakdash-5S. Fig. \ref{ris:M_vs_T_FeCoSi} gives the temperature scans of the magnetization for different compounds at the field $H = 100$ mT.   
The experimental magnetization curves were used to estimate the ordering temperatures $T_c$ as the position of the maximums at the derivative $dM/dT$ (Fig.\ref{ris:M_vs_T_FeCoSi}).

\begin{figure}[ht]
        \begin{minipage}{0.99\linewidth}
       \center{\includegraphics[width=1\linewidth]{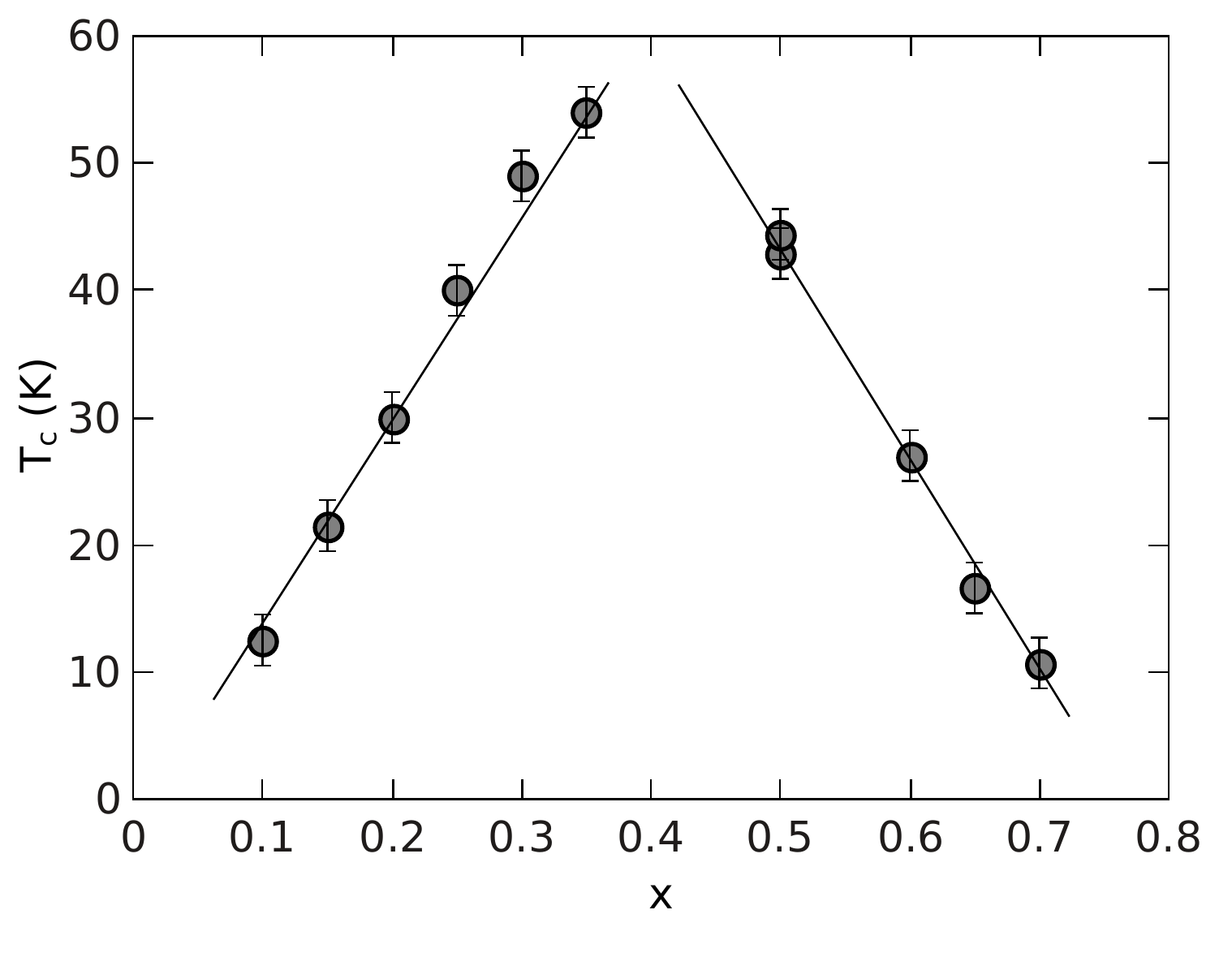}}
        \end{minipage}
        \caption{Dependence of the critical temperature $T_c$ on the concentration $x$ of Fe$_{1-x}$Co$_x$Si compounds.}
        \label{ris:T_dep_FeCoSi}
\end{figure}

The same analysis has been applied to the SQUID data for the samples studied in Ref.  \cite{Grigoriev_PRL_2009} for $x = 0.1 \div 0.5$. The  $x$-dependence of the critical temperature $T_c$ in the range $x = 0.1 \div 0.7$ is shown in Fig.\ref{ris:T_dep_FeCoSi}. $T_c$ increases  monotonically on increase of $x$ from $0.1 \div 0.4$. For $x > 0.4$: $T_c$ decreases again monotonically with $x$  and approaching  0 at $x \rightarrow 0.8$, proving that the compounds under study are magnetically ordered up to  $x = 0.7$. Notably, the pure compounds do not show any magnetic ordering while their solid solutions show a remarkable compositional dependence of the ordering temperature. If the exchange interactions is a function of the number of Fe-Co pairs, then for an ideal mixture the maximum of $T_c$ is expected at $x=0.5$; the experiment gives $x=0.4$ and the reason for the difference is still to be found.

%For a  helical magnetic structure absolute configuration is defined by the sense of corresponding magnetic spiral. The methodology of experimental determination of the magnetic chirality with small angle neutron diffraction is shown in Ref. \cite{Grigoriev_PRL_2009}. 

The chirality of the magnetic structure was determined using polarized neutron diffraction \cite{Maleev_PSS_1962, Blume_PR_1963}. We used the protocol similar to one described in Refs. \cite{Grigoriev_PRL_2009, Grigoriev_PRL_2013} for the data analysis. The polarized small-angle neutron scattering (SANS) was carried out at the SANS-1 instrument at the Meier-Leibniz-Zentrum in Garching. The wavelength of the neutron beam was set in the range from 0.6 nm to 1.2 nm depending on the needed resolution. A position sensitive detector with 128 $\times$ 128 pixels and a pixel size of 8 mm was used. These settings allowed us to cover a $Q$ range from 2 $\times$ 10$^{-2}$ to 1 nm$^{-1}$. The initial polarization of the neutron was $P_0 \approx$ 0.9.

\begin{figure}[ht]
        \begin{minipage}{0.99\linewidth}
        \center{\includegraphics[width=1\linewidth]{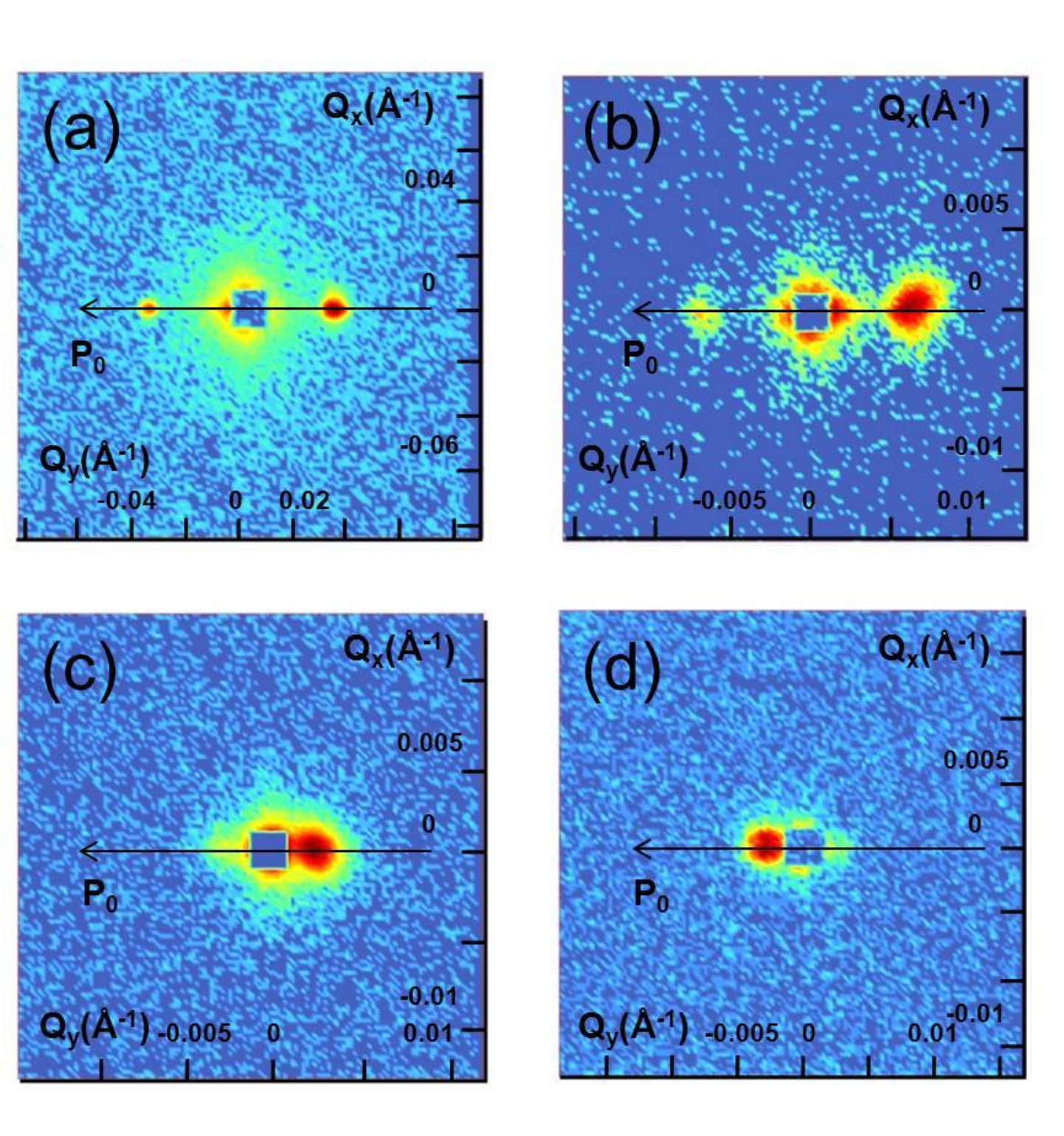}}
        \end{minipage}
        \caption{(color online). Maps of polarized SANS intensities of MnSi (a) and of Fe$_{1-x}$Co$_x$Si with $x = 0.5$ (b), 0.6 (c), 0.7 (d) for polarization $+\textbf{P}_0$ along the guide field at $T \approx 3.5 $K.}
        \label{ris:scatteringmaps_pol}
\end{figure}

Figure \ref{ris:scatteringmaps_pol} shows the polarized small angle neutron scattering maps for the compounds MnSi and Fe$_{1-x}$Co$_x$Si with $x$ = 0.5, 0.6 and 0.7 at low temperature. As one can see, the MnSi reference sample shows a maximum of the scattering intensity at the right part of the detector with an initial polarization of the neutron beam along the magnetic guide field [Fig. \ref{ris:scatteringmaps_pol}(a)]. For $x$ = 0.5 [Fig. \ref{ris:scatteringmaps_pol}(b)] and $x$ = 0.6 [Fig. \ref{ris:scatteringmaps_pol}(c)] the behaviour is similar to the MnSi reference sample and the maximum of the scattering intensity is at the right side of the scattering maps. In agreement with definitions given previously \cite{tanaka_JPSJ_1985, Grigoriev_PRL_2009}, the magnetic chirality for this configuration is $\gamma_m = -1$. Clearly, the Fe$_{0.3}$Co$_{0.7}$Si sample shows opposite behaviour [Fig. \ref{ris:scatteringmaps_pol}(d)] having $\gamma_m = +1$.

\begin{figure}[h]
        \begin{minipage}{0.99\linewidth}
        \center{\includegraphics[width=1\linewidth]{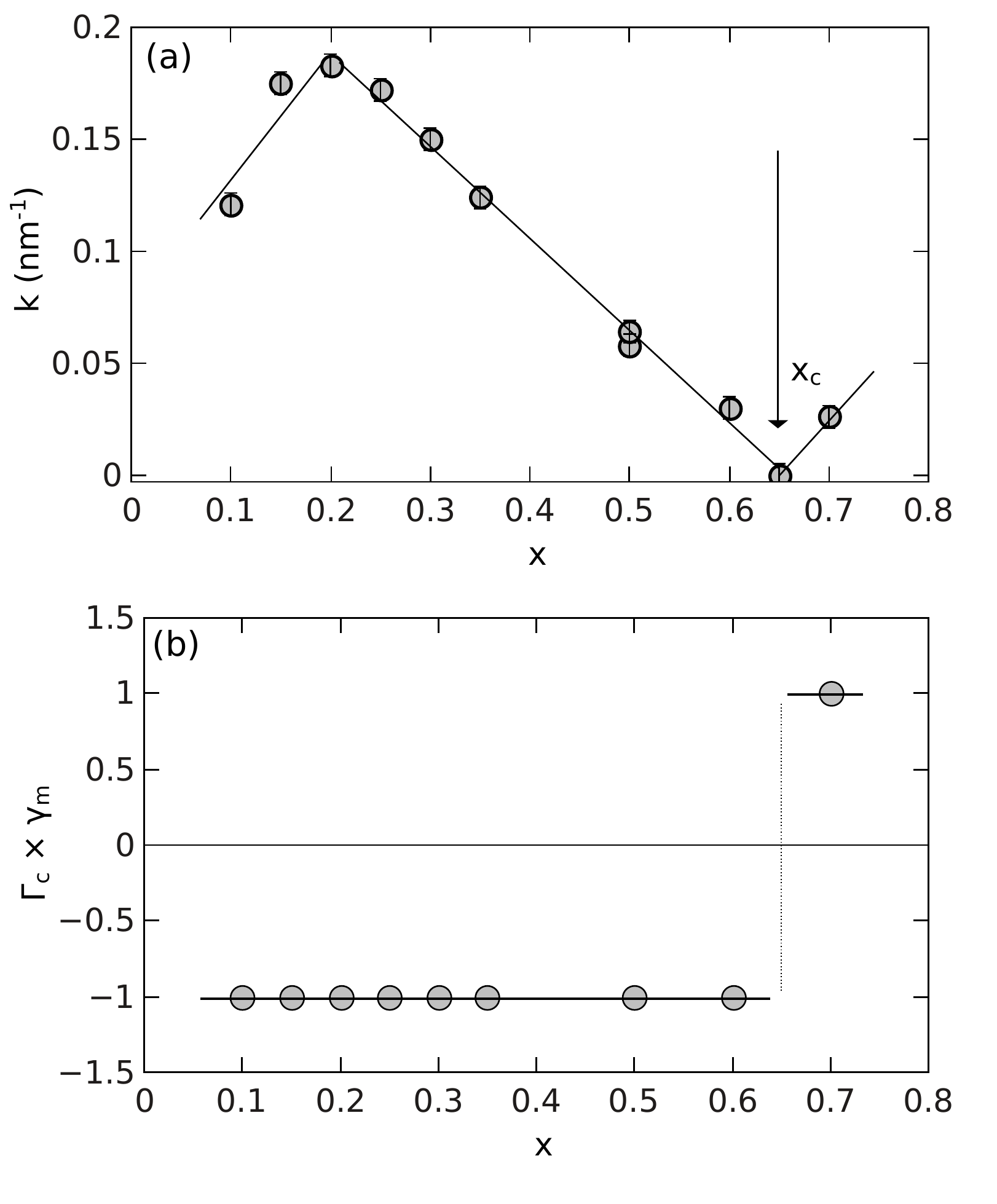}}
        \end{minipage}
        \caption{Dependence of (a) the helix wavevector $k$ and (b) the product of structural and magnetic chiralities $\Gamma_c \times \gamma_m$ on the concentration $x$. $x_{c}$ shows where $k$ goes to zero.}
        \label{ris:k-gamma_dep_FeCoSi}
\end{figure}

The helix wavevector $\vert k_s \vert$ has been extracted from the scattering maps at low temperature ($T \approx 3.5$ K). Figure \ref{ris:k-gamma_dep_FeCoSi}a shows the $x$ dependence of $\vert k_s \vert$, the product of the lattice chirality $\Gamma_c$ and the magnetic chirality $\gamma_m$ is shown in Fig \ref{ris:k-gamma_dep_FeCoSi}b. For $\vert k_s \vert$ the value increases from $\vert k_s \vert  = 0.121$ nm$^{-1}$ for $x = 0.1$ to a maximum of $\vert k_s \vert  = 0.185$ nm$^{-1}$ for $x = 0.2$. For $x > 0.2$ the value decreases to a minimum $\vert k_s \vert \rightarrow 0 $ nm$^{-1}$ at the critical concentration of $x_c = 0.65$ and increases again to $\vert k_s \vert  = 0.026$ nm$^{-1}$ for $x = 0.7$. The helix wavevector $k$ and the Dzyaloshinskii constant $D$ are linked via the equation:

\begin{equation}
 k = \frac{SD}{A},
\end{equation}
where $S$ is an average spin per unit cell and $A$ is the spin wave stiffness \cite{Maleyev_PRB_2006}. The spin wave stiffness and the spin value are expected to be monotonic functions of the Co content \cite{Grigoriev_PRB_2007_092407, Manyala_Nature_2000}, therefore $|k_s| \rightarrow 0$ implies that $|D| \rightarrow 0$ at $x_c$.

%As all Fe$_{1-x}$Co$_x$Si samples have a right handed crystal chirality ($\Gamma_c = +1$), then samples with a Co-concentration below the critical concentration of $ x \approx 0.65$ show a left handedness of the magnetic chirality (similar to the MnSi reference sample; $\gamma_m = -1$), while the sample with a concentration of $x = 0.7$ has opposite to this a right handed magnetic chirality ($\gamma_m = +1$). That means that the magnetic chirality is flipped with the change of the concentration in Fe$_{1-x}$Co$_x$Si compounds. 

The same concentration $x_c$ separates two regions with opposite values of the product $\Gamma_c \times \gamma_m$, therefore, $D$ not only goes through zero at $x_c$ but also changes its sign.
%Our findings are summarized in  Table \ref{tab:table2}.

To summarize, we show that the chiral magneto-lattice coupling mapped phenomenologically as the DMI could be applied to control magnetic chirality as needed for yet illusive spintronics applications.

The sign of the Dzyaloshinskii constant $D$ defines the chirality of magnetic helix relative the structural chirality. 
The product $\mathrm{sgn} (D) \times \Gamma_c \times \gamma_m$ is an invariant with respect to inversion and time-reversal operations ensuring that left-handed and right-handed polymorphs have the same energy. 
The sign of $D$ depends on 3d-element occupying the metal site in Fe$_{1-x}$Co$_x$Si, and also in monogermanides \cite{Grigoriev_PRB_2014,Tokura_2, Grigoriev_PRL_2013}.
The difference in the critical concentrations, $x_c = 0.65$ for monosilicide and $x_c = 0.6$ for monogermanides \cite{Grigoriev_PRB_2014} is rather small; more detailed sampling near the critical concentration has to be done to find whether this difference is significant. Interesting to note that the different systems Mn$_{1-x}$Fe$_x$Ge also shows similar behavior with a relatively close concentration $x_c = 0.75$.
Those findings together with a complex nature of the transformation of the helical magnetic structure to a ferromagnetic-like at $x \rightarrow x_c$  should be subject of further theoretical and experimental studies.

The work was supported by the Russian Foundation for Basic Research 
(grant No. 13-02-01468,
and 14-22-01073) and subsidy of the Ministry of Education and Science of 
the Russian Federation No. 14.616.21.0004.

%\bibliographystyle{apsrev}
%\bibliography{biblio}

\end{document}